\documentclass[preprint]{aastex}

\begin{document}

\title{Deep HST WFPC2 Photometry of M31's Thick Disk(?)}

\def\lea{\mathrel{<\kern-1.0em\lower0.9ex\hbox{$\sim$}}}
\def\gea{\mathrel{>\kern-1.0em\lower0.9ex\hbox{$\sim$}}}

\author{Ata Sarajedini\altaffilmark{1}}
\affil{Department of Astronomy, University of Florida, P. O. Box 
112055, Gainesville, FL  32611-2055}
\email{ata@astro.ufl.edu}

\author{Jeffrey Van Duyne\altaffilmark{2}}
\affil{Astronomy Department, Wesleyan University, Middletown, CT  06459}
\email{jeff@astro.wesleyan.edu}

\altaffiltext{1}{Guest User, Canadian Astronomy Data Centre, which is 
operated by the Dominion Astrophysical Observatory for the National 
Research Council of Canada's Herzberg Institute of Astrophysics.}

\altaffiltext{2}{Current Address: Department of Astronomy, Yale 
University, P. O. Box 208101, New Haven, CT 06520-8101}

\begin{abstract}
We present deep color-magnitude diagrams (CMDs) for a field along the 
outer disk of M31 based on archival Hubble Space Telescope Wide Field Planetary 
Camera 2 observations in the F555W ($\sim$V) and F814W ($\sim$I) 
filters. The CMDs, which contain a total of about 50,000 stars, feature a 
prominent red giant branch (RGB) along with a significant population of helium 
burning red clump stars. In addition, they exhibit the rarely seen asymptotic 
giant branch clump as well as a weak `Pop II' horizontal branch. There 
is also the hint of a $\sim$2 Gyr subgiant branch at the faintest 
levels of the CMDs. After adopting an M31 distance of $(m-M)_0 = 24.5$ 
and a reddening of $E(B-V) = 0.08$, we draw the following conclusions.
1) The I-band absolute magnitude of the helium burning red clump 
stars is $M_{I}(RC) = -0.29 \pm 0.05$, which is in accord with the 
value derived from Hipparcos parallaxes of solar neighborhood clump 
stars by Stanek \& Garnavich.
2) The metallicity distribution function constructed from 
bright RGB stars shows a characteristic shape; however, a
pure halo population consisting of metal-poor and
intermediate metallicity components (as advocated in the literature) are not 
sufficient to account for this shape. Instead, an additional
Gaussian component with $\langle$$[Fe/H]$$\rangle$$ = -0.22 \pm 0.26$,
comprising 70\% of the total number of stars, is required.
3) A comparison of our CMD with theoretical isochrones indicates that 
the majority of stars in our M31 field have ages that are $\gea$1.5 Gyr.  
4) These points, along with the physical location 
of our field in M31, suggest that we are observing the thick disk 
population of this galaxy.
\end{abstract}


\section{Introduction}

The determination of star formation histories for spiral galaxies 
is an important ingredient in constraining models of galaxy formation
(Bullock 1999; Grebel 2000; Silk 2000).  
The three most readily available subjects are, of
course, our own Milky Way galaxy, M31, and M33.  Being within the Milky Way
makes much of this work difficult, which is why a good understanding
of the disk and halo of M31 is so important.  However, little is known 
about the early star formation history of M31 due primarily to the lack of
sufficiently deep photometry of its disk and halo.

There is a rich history of ground-based work dealing with the stellar 
populations of M31. The reader is referred to the review by van den 
Bergh (1999) for a discussion of many of these results. For the purposes
of the present paper, we concentrate on previous photometric
investigations of the outer 
disk and inner halo. We begin by noting that ground-based photometry 
has only generated color-magnitude diagrams
(CMDs) of M31 field stars as deep as the horizontal branch, which is
roughly at I=24.5 (Durrell, Harris, \& Pritchet 1994). Nevertheless, 
a great deal can be learned about M31 by examining the colors 
and magnitudes of the brighter stars. 

Pritchet \& van den Bergh (1988) present 
photometry for an inner halo field located 40 arcmin southeast of the 
nucleus along the minor axis. They find a mean abundance of 
$[Fe/H] \sim -1.0$, similar to the Galactic globular NGC 6171, with a 
dispersion of $\sim$0.3 dex. The work of Davidge (1993) on an inner 
halo field situated on the opposite side of the galaxy
near NGC 205 confirms these results; namely, 
he finds a mean $[Fe/H]$ between --1.3 and --0.7 along 
with a metallicity dispersion of $\sim$0.3 dex. Davidge (1993) also 
presents observations that sample the outer disk of M31 at a location
25.6 arcmin from the nucleus. As one would expect, this field presents 
a rather complicated stellar population profile, suggesting a 
significant range in age and metallicity among the stars. Previously, 
Hodge \& Lee (1988; see also Hodge, Lee, \& Mateo 1988 and Massey, 
Armandroff, \& Conti 1986) imaged six disk fields in M31 and
were able to study reddening variations in their fields along with 
the properties of the luminosity functions. The color-magnitude 
diagrams published by Richer, Crabtree, \& Pritchet 
(1990) and Morris et al. (1994) reveal the metal-rich nature of the 
M31 disk. The latter study finds a mean metal abundance that is 
significantly higher than that of 47 Tuc ($[Fe/H] = -0.7$).
Recent wide field imaging observations of M31's outer disk  
by Cuillandre et al. (2001) are used to investigate the correlation 
between the young star population, HI column density, and the 
dust content.

Extending the ground-based studies with the Hubble Space Telescope 
(HST) has resulted in CMDs that reach 2--3 magnitudes fainter than the 
M31 field horizontal branch. In spite of this, HST has
been used relatively sparingly in examining the field halo and disk 
populations of M31 (Holland, Fahlman, \& Richer 1996; 
Rich, Mighell, \& Neill 1996).  Instead, the
majority of HST photometry of M31 has been directed at the star
clusters (Fusi Pecci et al. 1996; Holland et al. 1997;
Jablonka et al. 2000).

One of these HST programs (GO-5420) was designed to construct
CMDs of M31 star clusters using broad-band filter observations
with the Wide Field and Planetary Camera 2 (WFPC2). One object on 
the target list was intended to be the M31 globular cluster G272
($\alpha$$_{2000} = 00^h 44^m 14.5^s$, 
$\delta$$_{2000} = 41$$^{\circ}$ 19' 19.8'').
Apparently however, a bright nearby star was imaged instead
($\alpha$$_{2000} = 00^h 44^m 47.7^s$, 
$\delta$$_{2000} = 41$$^{\circ}$ 18' 48'').
As a result, the WFPC2 observations of this field were apparently
ignored and do not appear in any of the subsequent analysis papers
which focus mainly on the clusters (Fusi Pecci et al. 1996). 

Having come across these images in the HST archive and realizing that
the exposure times were extraordinarily long (Section 2), the
scientific utility of these observations became apparent. 
As discussed in Sec. 2, this region of M31 samples both the
disk and halo stellar populations. 
The next section describes the observations and data reduction.  
Section 3 presents the resulting CMDs 
and the properties of the field that can be determined therein, such
as the magnitude of the red clump and the metallicity distribution
function. This section also contains the results
of our artificial star experiments, further analysis of the
populations and their properties, and discussions 
of our results.  Section 4 presents our conclusions and suggestions
for future work.

\section{Observations and Data Reduction}

Our field center is situated at a $\alpha_{2000}$$ = 00^{h} 44^{m} 50.6^{s}$ 
and a $\delta_{2000}$$=+41^{o} 19' 11.1''$,
which is positioned at a $\sim$45\(^{o}\) angle from the minor
axis of the M31 disk, as seen in Fig. 1a.  Figure 1b is an expanded view
of this area, showing the HST-WFPC2 orientation on the sky.
A scale conversion from arcminutes to kpc at 
our adopted M31 distance of 790 kpc (Da Costa et al. 2000; Durrell, 
Harris, \& Pritchet 2001)
places our field at a projected distance of 5.5 kpc from the
center of M31.  

We obtained the processed images of the `G272' field 
from the Canadian Astronomy Data 
Centre\footnote{Canadian Astronomy Data Centre is operated by the 
Dominion Astrophysical Observatory for the National Research Council 
of Canada's Herzberg Institute of Astrophysics.}.
The observations were taken on 1995 January 23 and include two
frames taken in the F555W ($\sim$ V) filter and five taken
with the F814W ($\sim$ I) filter with total exposure times of 3800s 
and 10,800s, respectively.  The Wide Field 3 chip was
ignored due to the extreme saturation effects of the 
foreground star, as is clearly seen in Fig. 1c.  Table 1 provides
more detailed information about the observations.

The photometric reduction was performed on the three remaining
chips (PC1, WF2, and WF4), in a manner similar to the procedure described in
Sarajedini et al. (2000), utilizing the DAOPHOT II / ALLSTAR / ALLFRAME 
(Stetson 1994) profile fitting software. The reader is referred to
that paper for details. In summary, high single-to-noise WFPC2 
F555W and F814W point spread functions (PSFs), kindly provided by 
Peter Stetson, were fitted to all detected profiles on each 
individual image using the ALLFRAME software. The resultant instrumental
magnitudes were edited and matched to form colors. Aperture corrections
were applied as described in the next paragraph to bring the
total magnitudes to a 0.5'' radius. Then, a correction for the
well-known charge transfer efficiency (CTE) problem was applied
(Sarajedini et al. 2000) and standard magnitudes were calculated using
the equations of Silbermann et al. (1996). These describe the photometric
system established by the HST ``Cepheid Distance Scale" key project and
are coupled to the PSFs we used earlier in the reduction procedure.

The only departure from the procedure
adopted by Sarajedini et al. (2000) was in the determination of the 
aperture corrections for the PC1 chip. Because of the lack of sufficient 
numbers of bright stars in the Planetary Camera, we adopt the 
following procedure to ensure consistency between the photometric scales 
of the three CCDs. We derived and applied aperture corrections to the
WF2 and WF4 data as described by Sarajedini et al. (2000). Then, under
the assumption that the spatially adjacent regions of the three chips
feature the same peak magnitude for the red clump (see Figs. 1 and 2), we
offset the photometry from each chip to match the average peak red clump
magnitude of the WF photometry. The offsets to the WF data
(0.03 mag in V and 0.04 in I) correct systematic offsets between
the photometric scales of these two CCDs and the offsets to the PC1
data (0.03 mag in V and I) account for the fact that no
aperture corrections were applied. In any case, after the application
of these offsets, we estimate that the photometric scales of the 
three CCDs are consistent to within $\pm0.015$ mag.

\section{Results and Discussion}

\subsection{Color-Magnitude Diagrams}

The left panels of Figures 2a through 2c show the color-magnitude
diagrams (CMDs) of the PC1 (5249 stars), WF2 (22793 stars), and WF4 
(20943 stars) fields, respectively, in the
apparent $(V,V-I)$ plane, while the right panels show the same fields 
in the $(I,V-I)$ plane. 
The two most prominent features in these CMDs are
the well-populated red giant branch (RGB) which exhibits a
large range in color and an obvious red clump at V$\sim$25.5
and I$\sim$24.3.
There is also a hint of a `Pop II' horizontal branch located 
blueward of the red clump extending from 
$V-I$$\sim$0.1 to $V-I$$\sim$0.7 with $V$$\sim$25.4 and appearing
most prominently in the WF2 CMD. This 
feature is likely to be associated with the relatively small
metal-poor population present in our field as discussed in Sec. 3.6. 
The tip of the first ascent RGB appears to be at I$\sim$21.8. All of
the CMDs also feature a main sequence (MS) of young stars with ages 
between $10^{8}$ and $10^{9}$ years (see below).  
Note also the existence of what appears to be a subgiant branch
developing at $V\sim$26.5 (see Fig. 12 and the discussion in Sec. 3.7). 
Lastly, there is a
clump of stars at $V$$\sim$24.7 and $I$$\sim$23.2,
which we tentatively identify with the asymptotic giant branch (AGB)
clump (Gallart 1998).

To investigate radial variations in the CMDs, we divided our sample
into three regions (1, 2, and 3) based on radial distance from the 
center of M31. Each star's right ascension and declination
was determined using the IRAF\footnote{IRAF is distributed by 
the National Optical Astronomy Observatories which are operated 
by the Association of Universities for Research in Astronomy, Inc. 
under cooperative agreement with the National Science Foundation.}
routine METRIC and then separated into three 1 arcmin (projected) wide
regions. The black lines in Fig. 1c show these divisions; region 1, 
which is closest to the M31 nucleus, is on the right hand side of this figure.

\subsection{Distance and Reddening}

The reddening of our M31 field was estimated using the Burstein \& Heiles 
(1982) maps. At the location of our field ($l = 121.60, b=-21.53$),
these maps indicate a reddening of $E(B-V) = 0.08$, which is consistent
with the value used by Holland et al (1999).  We adopt the
relations $E(V-I) = 1.25 E(B-V)$  and $A_{I} = 1.48E(V-I)$ from
Schlegel et al. (1998) for the HST filters utilized herein.

It is important to point out at this juncture that we have 
deliberately neglected the effects of differential reddening across 
our field and along the line of sight. Differential reddening due to
the Galactic foreground across our (small) WFPC2 field of view is likely to be 
negligible. However, the effects of dust internal to M31 may have a 
significant effect on our results. We return to this point in Sec. 3.6.

For the distance modulus of M31, we adopt a value of $(m-M)_0 = 24.5 \pm 0.1$ 
based on the mean of those presented by Da Costa et al. (2000), 
who quote distances to M31 based on the field halo RR Lyraes and 
giant stars as well as the M31 globular clusters. This distance
modulus is on the scale of Da Costa \& Armandroff (1990).
Figures 3a, b, and c depict our distance and reddening corrected CMDs for
the three radial regions defined above.
We note that the appearance of the three CMDs is qualitatively 
indistinguishable. 

A particularly striking feature is that 
the tip of the RGB, at $M_{I}\sim -3$, is significantly fainter 
in all three CMDs 
than the canonical value of \(M_{I} 
= -4.05\pm 0.10\) (Da Costa \& Armandroff 1990, hereafter DCA; Sakai
et al. 2000; Bellazzini, Ferraro, \& Pancino 2001).  In contrast, 
the RGB tip is easily identified at 
I$\sim$20.6 ($M_{I}$$\sim$--4.0) in Fig. 2 of Holland et al. (1996)
showing the CMD of the field around the M31 globular cluster G302 constructed 
from HST 
observations similar to those considered herein. The faintness of the RGB tip
is consistent with the fiducials of very metal rich star
clusters such as NGC 6553 and NGC 6528 (Bica, Barbuy, \& Ortolani 1991), 
giving us our first hint that this field in M31 contains a very
metal-rich stellar population. More discussion of this phenomenon is
provided in Sec. 3.4.

\subsection{Horizontal Branch and Red Clump}

As mentioned above, the most conspicuous feature of the M31
field star CMD presented herein is the helium burning red clump (RC).
Based on Fig. 3, we find peak values of $M_{I}(RC)=-0.28 \pm 0.05$,
$-0.30 \pm 0.05$, and $-0.29 \pm 0.05$ for regions 1, 2, and 3, 
respectively. All of these values are identical to within the errors
and very close to the red clump absolute magnitude advocated by 
Stanek \& Garnavich (1998); they find $M_{I}(RC)=-0.23 \pm 0.03$ based on
Hipparcos parallaxes of solar neighborhood red clump stars.  

Not so conspicuous but nevertheless present in the CMDs 
is a `Pop II' horizontal branch most easily seen in the CMD of 
Region 1 shown in Fig. 3a. The location of this feature is evident
in Fig. 4 where we have superimposed the HBs and RGBs of the well-known
Galactic globular clusters M68 (Walker 1994) and M5 (Johnson \&
Bolte 1998). The metallicities, distance moduli, and reddenings 
for these clusters are taken from Table 1 of Layden \& Sarajedini 
(1997) adopting $A_{V} = 3.1E(B-V)$. 

\subsection{Asymptotic Giant Branch}

As already noted, the CMDs of the G272 field show evidence for the AGB 
clump. Analogous to the RGB clump (Fusi Pecci et al. 1990; 
Sarajedini \& Forrester 1995; Ferraro et al. 1999), the AGB feature is caused 
by a temporary pause in the evolution of stars as they proceed up the 
AGB and is related to the formation of the Helium burning shell 
(Caputo et al. 1989). As shown by Alves \& Sarajedini (1999), the 
luminosity of the 
AGB clump (alternatively referred to as the AGB bump because of its 
appearance in a luminosity function) is primarily a function of the 
mean age and metal abundance of 
the stellar population. To eliminate the effects of distance 
uncertainties, the magnitude of the AGB bump is measured relative to the 
horizontal branch/red clump [$\Delta$$V(AGBC-RC)$]. 

Building upon the work of Castellani, Chieffi, \& Pulone (1991) and
Ferraro (1992), Alves \& Sarajedini (1999) 
present $\Delta$$V(AGBC-RC)$ values for four Galactic globular 
clusters - M5, NGC 1261, NGC 2808, and 47 Tuc, the only ones in which 
the AGBC could be isolated at the time.
They showed that the predictions of the theoretical 
models are in good agreement with the ages, metallicities, and
$\Delta$$V(AGBC-RC)$ values of these clusters. 

For our M31 field, we have already measured the luminosity of the red 
clump and thus adopt $M_{I}(RC) = -0.29 \pm 0.05$ as the mean of the 
three regions shown in Fig. 3. The average color of the red clump 
is $(V-I)_{o} = 1.00 \pm 0.05$ making the absolute V magnitude of 
the red clump equal to $M_{V}(RC) = 0.71 \pm 0.07$. Performing the 
same calculation on the AGB clump stars, we find 
$M_{V}(AGBC) = -0.10 \pm 0.07$ leading to a difference of 
$\Delta$$V(AGBC-RC) = -0.81 \pm 0.10$. Assuming that the G272 field is not 
significantly older than the Galactic globulars, Fig. 6 of Alves \& 
Sarajedini (1999) suggests that the metal abundance of this field is 
likely to be greater than $\sim$--0.4 dex. However, because the 
Alves \& Sarajedini models do not extend to more metal-rich regimes,
it is difficult to say anything more definitive. In any case, we have 
once again confirmed the metal-rich nature of the stellar population
in the `G272' field.

\subsection{Artificial Star Experiments}

As noted above, the RGB shows a rather large color width; one of the
factors that could influence this width is the photometric error. 
In order to assess its importance, we performed a number 
of artificial star experiments.
These consist of placing stars of known magnitude and color on our
original CCD frames and applying the same reduction techniques as
those used in the reduction of the genuine stars.
We placed these stars on each frame arranged in a grid pattern
under the constraint that no two artificial stars were within 2
PSF radii of each other. The resultant images were reduced using the 
same method described in Section 2.  Four trials were performed  
with a total of 1668 stars recovered after reducing the three CCDs.  
The magnitudes of the artificial stars were taken from a locus 
with \(22.55 < I < 24.95\) along the RGB as displayed in Fig. 5a.
The locations of the stars at their recovered magnitudes and
colors are shown by the points in Fig. 5a. The difference between
the input and measured colors is shown in Fig. 5b as a function
of the apparent I magnitude. 

To determine the photometric error along the RGB, we
fit a Gaussian curve to the color distribution of artificial 
with $22.55 \leq I \leq 22.75$ (Fig. 5c). As discussed in the next section, we
have chosen to use the brighter RGB stars to study the metallicity
distribution of this field. Thus,
at I = 22.65, which corresponds to $M_{I} = -2$ using our adopted 
distance modulus of $(m-M)_{I}= 24.65$, the artificial stars exhibit a standard
deviation of \(\sigma_{err}=0.049\) in (V--I). This quantity is taken to be the 
mean photometric error at this magnitude (i.e. I=22.65, $M_I = -2$)
and will be used in the metal abundance analysis of the next section.

The lower panel of Fig. 6 shows our actual photometry in the distance 
and reddening corrected CMD and the solid lines are the RGBs of 
M15 (left) and NGC 6791 (right), which encompass the majority of RGB 
stars; the dotted lines enclose the magnitude 
range $-2.1 \leq M_{I} \leq -1.9$ ($22.55 \leq I \leq 22.75$) over 
which the color histogram shown in the upper panel has been 
constructed. The solid line in the upper panel shows the weighted least
squares fit of a 
Gaussian function to this distribution. The standard deviation of 
this fit is $\sigma_{obs}$ = 0.182, which is much larger than the 
contribution purely from the photometric error. In fact, if we 
subtract $\sigma_{err}$ from $\sigma_{obs}$ in quadrature, we find an 
intrinsic color spread of $\sigma_{int}$ = 0.175 mag which is represented 
by the dashed curve in the upper panel of Fig. 6. The solid and 
dashed curves have been scaled to have the same area. Note that these 
two distributions are virtually indistinguishable suggesting that the 
influence of photometric error on the metallicity distribution 
function is likely to be negligible.

\subsection{Metallicity Distribution Function}

Now that we have utilized artificial star experiments to quantify the 
photometric errors on the RGB, it is possible to use the color of each 
RGB star to construct a metallicity distribution function (MDF). 
Figures 3a through c show our distance and reddening corrected CMDs 
compared with the RGB fiducials of the Galactic globular clusters 
M15, 47 Tuc (both from DCA), and NGC 6553 (Guarnieri et al. 1998), 
along with the RGB of the open cluster NGC 6791 (Garnavich et al. 1994). 

We prefer to use observed cluster fiducials over theoretically 
calculated RGB sequences because of the uncertainties in the 
color-temperature calibration and our limited knowledge of the 
physics of convection. We attempted to make use of the empirical 
RGB grid constructed by Saviane et al. (2000), but realized that the 
metallicity range of the grid is not sufficient to cover the range of 
stars observed in our M31 field. In any case, over the metallicity 
range common to both RGB sets, there is good agreement between the 
Saviane et al. (2000) grid and our RGBs discussed below.

The adopted distance moduli and reddenings used to place the cluster 
fiducials in Fig. 3 are given in Table 2. The 
latter values are taken predominantly from the work of DCA. 
In the case of NGC 6553, we adopted the reddening advocated by 
Guarnieri et al. (1998) based on their application of the simultaneous 
reddening and metallicity method developed by Sarajedini (1994). For 
NGC 6791, we used an average of the values quoted by Garnavich et al. 
(1994) and Chaboyer, Green, \& Liebert (1999). 
The distance moduli in 
Table 2 deserve a more detailed explanation. This is because we utilized 
a slightly different technique for clusters with RR Lyrae variables 
as compared to those with red clumps (RC). To begin with, we note again
that our adopted M31 distance 
modulus is that of Da Costa et al. (2000) and is based on 
$M_{V}(RR) = 0.17 [Fe/H] + 0.82$ (Lee, Demarque, \& Zinn 1990) for 
the RR Lyraes. For the three clusters with red clumps, we 
must modify the calculated 
$M_{V}(RR)$ to take account for the fact that the red clump and the 
RR Lyraes have different absolute magnitudes and that this magnitude
is a function of metallicity and age (Cole 1998; Alves \& Sarajedini 
1999; Sarajedini 1999; Girardi \& Salaris 2001). We consider each 
of the red clump clusters in turn.

\noindent {\it 47 Tuc:} We make use of the results published by Sarajedini, 
Lee, \& Lee (1995, hereafter SLL) which are based on synthetic HB models 
that are consistent with our distance scale. From the work of SLL, 
we see that $M_{V}(RC) = 0.64$ so that $(m-M)_{V} = 13.42$ and 
thus $(m-M)_{I} = 13.37$ for 
47 Tuc. This is somewhat smaller than the DCA value of $(m-M)_{V} = 
13.51$ which was based on arbitrarily setting the apparent RR Lyrae 
magnitude of 47 Tuc 0.15 mag fainter than the red clump magnitude. 
The globular cluster compilation of Harris (1996) gives 
$(m-M)_{V} = 13.37$.

\noindent {\it NGC 6553:} Since the age of NGC 6553 is similar to that 
of 47 Tuc (Zoccali et al. 2001), we need only to correct the red 
clump luminosity of 47 Tuc for the metallicity difference between 
these two clusters. This correction gives $M_{V}(RC) = 0.71$ so that 
$(m-M)_{V} = 16.17$ and $(m-M)_{I} = 15.18$ for NGC 6553. Our values 
compare favorably with the distance modulus estimated by Guarnieri et 
al. (1998) of $(m-M)_{V} = 15.98 \pm 0.15$, but differs somewhat from
two of the most recent determinations; Zoccali et al. (2001) find  
$(m-M)_{V} = 15.70 \pm 0.13$ while Beaulieu et al. (2001) derive
$(m-M)_{V} = 15.4$.

\noindent {\it NGC 6791:} The metal abundance of NGC 6791 is greater 
than the most metal-rich models presented by SLL. As a result, the 
corrections for age and metallicity will be performed relative to 
the $M_{V}(RC)$ value of 47 Tuc using the HB models of Girardi et al. 
(2000, see also Crowl et al. 2001). Thus, because NGC 6791 is 0.99 
dex more metal-rich than 47 Tuc and 4 Gyr younger (Chaboyer et al. 
1999), we calculate its red clump to be at $M_{V}(RC) = 0.76$ making 
its apparent distance moduli equal to $(m-M)_{V} = 13.79$ and 
$(m-M)_{I} = 13.63$. As a comparison, we note that Garnavich et al. 
(1994) used a modulus of $(m-M)_{V}$$\sim$ 13.6 in their study of 
NGC 6791. 

The cluster RGBs shown in Fig. 3 are used to calibrate the dereddened 
RGB color at $M_{I}=-2$ [$(V-I)_{o,-2}$] as a function
of $[Fe/H]$. This magnitude level was chosen to minimize the effects 
of asymptotic giant branch stars while at the same time maximizing the 
effects of metallicity on color. The weighted least squares relation 
shown in the lower panel of Fig. 7 is given by

\begin{equation}
[Fe/H] = -27.24 + 46.18(V-I)_{0} - 26.49(V-I)_{0}^{2} + 5.16(V-I)_{0}^{3}.
\end{equation}

\noindent The root mean square deviation of the points from the relation is 
0.03 dex. The upper panel of Fig. 7 shows the metallicity dispersion
introduced by the color error of $\sigma_{err}=0.049$ at $M_{I}=-2$. 
This figure indicates that the photometric error translates to a typical 
error of only $\sim$0.2 dex in metallicity. Furthermore, for 
NGC 6553, which has the most uncertain distance modulus among 
the clusters in Table 2, an uncertainty of 0.2 mag in $(m-M)_{I}$
also leads to a metallicity error of $\sim$0.2 dex.

To construct the MDF, we take the dereddened color of each star within 
$\pm$0.1 mag of $M_{I}=-2$ and convert it to a metal abundance using 
Equation 1. In addition, the photometric error is converted to 
$\sigma_{[Fe/H]}$. A generalized histogram of the metallicities is then 
constructed by adding up the unit Gaussians representing the abundance 
of each star. The resultant MDF of the 271 stars in this magnitude 
range is shown in the upper panel of Fig. 8 wherein the filled circles 
and the solid line are the binned and generalized histograms, 
respectively. This comparison helps to illustrate which features are 
significant and which are diluted by the photometric errors. In this 
regard, we see that the MDF displays a prominent peak at 
$[Fe/H]$$\sim$ --0.1, a possible secondary peak at $[Fe/H]$$\sim$ 
--0.7, and an extended tail to the metal-poor regime. The lower 
panel of Fig. 8 displays the effect on the MDF of changing the adopted 
reddening by $\pm$0.02 mag in E(B--V). We note that the location of 
the peak changes by less than 0.1 dex and the overall shape 
of the MDF remains largely unchanged. 

We pointed out earlier that we have neglected differential reddening 
along the line of sight caused by dust internal to M31. 
Taking out this effect will tend to 
reduce the range of metallicities present in our MDF and it could
systematically lower the higher metallicity measurements. Because it is 
difficult to precisely account for this, the reader should keep this 
possibility in mind as the results of the analysis are presented.

\subsection{Comparison With Other MDFs}

Figure 9 shows the MDF for our M31 field (G272) located at a projected
radial distance of 23.9 arcmin compared with the M31 field halo MDFs from 
Holland et al. (1996, G302), Durrell et al. (1994), and Durrell et 
al. (2001). These are located at projected radial distances of 
32 arcmin, 40 arcmin, and 90 arcmin, respectively and have 
been scaled to have the same area as the `G272' field. In addition, we 
have shifted the Durrell et al. (2001) MDF by --0.3 in metal 
abundance to account for the fact that they quote $[M/H]$ rather than 
$[Fe/H]$ (Durrell 2001, private communication). 

There are a number of features to note in Fig. 9. First, 
all of the MDFs share the same overall shape; they feature a 
prominent peak at the metal-rich end with an extended tail to more 
metal-poor regimes. Relative to the peak, this tail appears to be 
most prominent in the Holland et al. (1996) and Durrell et al. (2001)
MDFs and less so in the Durrell et al. (1994) distribution. Second,
the metallicity of the `G272' peak occurs at $[Fe/H] \sim -0.1$, which is 
significantly more metal-rich that those of the other MDFs 
(see also Mould \& Kristian 1986; Pritchet \& van den Bergh 1988). 

Another way in which we can intercompare these MDFs is to scale 
them so that their metal-poor tails match, as shown in the bottom 
panel of Fig. 9. Keeping in mind that the three dashed MDFs are 
predominantly halo stars in M31, we find that 
the peak of the halo MDF is consistent with the `secondary peak' at 
$[Fe/H]$$\sim$ --0.7 in the `G272' distribution. This suggests that the 
latter population contains not just halo stars but also a significant 
population from another component. We tentatively assign this 
population to the thick disk of M31, an assertion that is not unrealistic 
judging from the physical location of our field
in M31 (Fig. 1a). Furthermore, we 
note that, because M31 halo populations do not exhibit a radial 
abundance gradient for R $\gea$ 5 kpc (van den Bergh 1999), it is 
unlikely to be the case that the dominant population in the `G272' 
field is simply a higher metallicity halo. Additionally, based on the 
discussion presented by Durrell et al. (2001), there {\it is} the possibilty 
that the metal-rich component in our field is actually the bulge of M31. 
However, given that the central regions of the bulge are around Solar 
metallicity (Renzini 1999; Jablonka et al. 1999; 2000) and that the 
mean abundance is expected to decrease at the rate of $\sim$0.1 
dex/kpc (Durrell et al. 2001), we should expect a peak 
bulge metallicity of $\sim$--0.6 dex 
at the location of our field. If present, this population would be 
indistinguishable from our `secondary peak' at $[Fe/H]$$\sim$ --0.7.

We can quantify our claim that the metal-rich peak in our MDF
belongs to the M31 thick disk by fitting multiple Gaussian 
distributions to the `G272' MDF representing the three populations 
that appear to be present - metal-poor and intermediate metallicity 
components that belong to the M31 halo and a metal-rich component that 
may be the thick disk. For the first two populations, we adopt the 
Gaussian parameters in Table 3 of Durrell et al. (2001). We can then 
fit a third Gaussian to our MDF to solve for the metallicity parameters
(peak and width) of the M31 thick disk. The two panels of Fig. 10 
illustrate the 
results of this exercise with the upper panel showing the fit to the 
binned histogram and the lower panel displaying the fit to the 
generalized histogram (see Fig. 8). There is relatively good agreement 
between the fitted data and the fits as well as between the fits to the 
binned and generalized histograms. This procedure suggests a mean metallicity 
of $\langle$$[Fe/H]$$\rangle$$ = -0.22 \pm 0.26$ for the M31 thick 
disk comprising 70\% of the stars in this field. This is in contrast with 
the abundances of the metal-poor and intermediate metallicity peaks of 
$[Fe/H] = -1.50 \pm 0.45$ (10\%) and
$[Fe/H] = -0.82 \pm 0.20$ (20\%), respectively, from Table 3 of Durrell et 
al. (2001) adjusted by --0.3 dex.

At this point,we return momentarily to the metallicities of the HB and red clump 
populations. We speculate that the most metal-poor 
component with $[Fe/H] \sim -1.5$ is likely to be associated with the 
above-mentioned `Pop II' HB while the most metal-rich population with 
$[Fe/H] \sim -0.2$ is probably producing the strong red clump. However,
the location of the helium burning stars associated with the intermediate 
metallicity ($[Fe/H] \sim -0.8$) component is unclear.

In Fig. 11, we compare the `G272' MDF with those of the Milky Way's 
thick disk stars (Wyse \& Gilmore 1995), field halo 
stars (Ryan \& Norris 1991), globular clusters (Harris 1996), and
M31's globular clusters (Barmby et al. 
2000). Interestingly, the `G272' field resembles the Milky Way's 
thick disk stars more than it does 
any of the other halo components (e.g. M31 globulars). One
interpretation of this would be that the stellar population of the `G272' 
field is dominated by the thick disk of M31, an assertion which would 
support our conclusion based on the comparisons in Fig. 9 above. However, 
this is a spurious line of reasoning because if we follow it 
to its logical end, it implies that the Durrell et al. (2001) MDF (of 
the M31 outer halo) does not represent the M31 halo because it 
does not resemble the MDF of the Milky Way halo field stars in Fig. 11b.

\subsection{Age of the Field Population}

Further evidence of our assertion that the `G272' field is dominated 
by M31 thick disk stars can be obtained by examining the age structure 
of this field.
Figure 12 shows the same CMDs as Fig. 3 with solar abundance (Z=0.019)
isochrones of $10^{8}$, $6.3 \times 10^{8}$, $10^{9}$, and 
$1.6 \times 10^{9}$ years 
(Girardi et al. 2000) overplotted. From these comparisons, we note
that the fraction of the stellar population younger 
than $10^{9}$ years is very small, comparable with the 
appearance of the G302 field CMD in Holland et al. (1996). In contrast, 
the observations of 
M31's disk presented by Williams \& Hodge (2001) reveal a 
significantly larger 
population of stars with ages younger than $10^{9}$ years. This difference 
suggests that the contribution of the M31 thin disk to the `G272' field is 
minimal. This, coupled with the fact that the dominant 
population in the `G272' field is likely to be $\gea$1.5 Gyr old, 
provides further circumstantial evidence that the `G272' field
is probably dominated by intermediate-to-old age thick disk stars.

\section{Summary and Conclusions}

We present the deepest HST/WFPC2 photometry of a field in M31. The 
VI color-magnitude diagram is based on 3,800s of exposure time 
in the F555W filter and 10,800s in the F814W filter. After adopting a 
distance of $(m-M)_0 = 24.5$ and a reddening of $E(B-V) = 0.08$, we draw the 
following conclusions.

\noindent 1) The I-band absolute magnitude of the helium burning red clump 
stars is $M_{I}(RC) = -0.29 \pm 0.05$, which is in accord with the 
value derived from Hipparcos parallaxes of solar neighborhood clump 
stars by Stanek \& Garnavich (1998).

\noindent 2) The V-band absolute magnitude of the asymptotic giant branch 
(AGB) clump stars is $M_{V}(AGB) = -0.10 \pm 0.07$; coupled with the 
red clump luminosity, this value is consistent 
with those predicted by the models of Alves \& Sarajedini (1999) 
for an intermediate age metal-rich population.

\noindent 3) The metallicity distribution function constructed from 
bright RGB stars shows a characteristic shape with a prominent peak 
at $[Fe/H] \sim -0.1$ and an extensive tail to metal poor regimes as low as 
$[Fe/H] \sim -2.5$. 

\noindent 4) A pure halo population consisting of metal-poor and
intermediate metallicity components (Durrell et al. 2001) is not 
sufficient to account for the shape of our MDF. Instead, an additional
Gaussian component with $\langle$$[Fe/H]$$\rangle$$ = -0.22 \pm 0.26$,
comprising 70\% of the total number of stars, is required.

\noindent 5) A comparison of our CMD with the theoretical isochrones 
of Girardi et al. (2000) indicates that the majority of stars in our 
M31 field have ages older than $\sim$1.5 Gyr.  

\noindent 6) All of the above points along with the physical location 
of our field in M31 suggest that we have observed the thick disk 
population of this galaxy.

\noindent We close by emphasizing the need for a robust model 
describing the spatial 
distribution of the various M31 components (e.g. thin disk, thick disk, 
bulge, halo). It is difficult for us to draw more compelling 
conclusions about the stellar populations of M31 without such a model.

\acknowledgements

We are grateful to Pat Durrell, Rupali Chandar, Andy Stephens, and
Pascale Jablonka for helpful comments on an early 
version of this manuscript. The suggestions of an anonymous referee 
also helped to clarify the presentation tremendously.
This project has benefited from financial 
support from NSF CAREER grant No. AST-0094048.

\break

\begin{deluxetable}{l c c}
\tablewidth{2.8in}
\tablecaption{Observing Log}
\tablehead{
\colhead{Dataset} & \colhead{Filter} & \colhead{Exp.time (sec)}}
\tablecolumns{3}
\startdata
u2gv0401 & F555W & 1500\\
u2gv0402 & F555W & 2300\\
u2gv0403 & F814W & 2300\\
u2gv0404 & F814W & 2300\\
u2gv0405 & F814W & 2300\\
u2gv0406 & F814W & 2300\\
u2gv0407 & F814W & 1600\\
\enddata
\end{deluxetable}

\begin{deluxetable}{lcccc}
\tablewidth{4.2in}
\tablecaption{Adopted Cluster Parameters} 
\tablehead{
\colhead{Cluster} & \colhead{$[Fe/H]$} & \colhead{$E(V-I)$} & 
\colhead{V(HB)\tablenotemark{a}} & \colhead{$(m-M)_I$}}
\tablecolumns{5}
\startdata
47 Tuc   & $-0.71\pm0.07$ & 0.05 & 14.06 & 13.37 \\
NGC 1851 & $-1.29\pm0.07$ & 0.03 & 16.05 & 15.43 \\
M2       & $-1.58\pm0.06$ & 0.03 & 16.05 & 15.48 \\
NGC 6397 & $-1.91\pm0.14$ & 0.23 & 12.90 & 12.18 \\
M15      & $-2.17\pm0.07$ & 0.13 & 15.86 & 15.29\\
\\[0.1cm]
NGC 6553 & $-0.28\pm0.15$ & 0.99 & 16.88 & 15.18 \\
NGC 6791 & $+0.28\pm0.18$ & 0.16 & 14.55 & 13.63 \\
\enddata
\tablenotetext{a}{All values from Da Costa \& Armandroff 1990
except for NGC 6553 (Guarnieri et al. 1998) and NGC 6791
(Sarajedini 1999).}
\end{deluxetable}

\clearpage

\figcaption{(a) The location of our HST/WFPC2 field is indicated by the 
square in this 1.5 x 1.5 degree digitized sky survey image of M31. 
North is up and east is to the left.  (b) Close-up view of the `G272'
field with the WFPC2 footprint overlaid. (c) The mosaicked WFPC2 
image with the `compass' markings showing the direction of north 
(arrow) and east. The black lines show the three radial regions
into which the photometry has been divided; region 1, which is
closest to the M31 nucleus, is on the right hand side.}

\figcaption{(a) The left panel shows the (V,V--I) color-magnitude 
diagram (CMD) of the PC1; the right panel shows the (I,V--I) CMD.
(b) Same as (a) except that the WF2 CMDs are shown. (c) Same as (a) 
except that the WF4 CMDs are shown.}

\figcaption{(a) The left panel shows the distance and reddening 
corrected $[M_V,(V-I)_{o}]$ color-magnitude diagram (CMD) for Region 
1 assuming $(m-M)_0 = 24.5$ and $E(B-V) = 0.08$; 
the right panel shows the $[M_I,(V-I)_{o}]$ CMD. The solid lines are 
the red giant branches of M15, 47 Tuc, NGC 6553, and NGC 6791.
(b) Same as (a) except that the Region 2 CMDs are shown. (c) Same as (a) 
except that the Region 3 CMDs are shown.}

\figcaption{The left panel shows the distance and reddening 
corrected $[M_V,(V-I)_{o}]$ color-magnitude diagram (CMD) for Region 1; 
the right panel shows the $[M_I,(V-I)_{o}]$ CMD. The solid lines are 
the red giant and horizontal branches of M68 and the dashed lines are those 
of M5. This comparison is designed to illustrate the presence of a Pop 
II horizontal branch in our M31 field.}

\figcaption{The results of the artificial star experiments are displayed.
(a) The solid sequence represents the red giant branch of 
artificial stars  placed on the images while the points show the 
locations of stars after they have been measured using our 
photometric technique. (b) The open circles illustrate the 
difference between the measured and actual V--I colors [$\Delta(V-I)$] 
of the artificial stars as a function of I magnitude. (c) For all artificial 
stars in the magnitude range $22.55 \leq I \leq 22.75$, we show the 
histogram of $\Delta(V-I)$ values with the fitted Gaussian 
distribution as the solid line. The standard deviation of this 
Gaussian is 0.049 mag, which we adopt as the photometric error in this 
magnitude range.}

\figcaption{The lower panel shows the distance and reddening corrected 
CMD of our M31 field along with the red giant branches of M15 
(left-most) and NGC 
6791 shown as the solid lines. The two horizontal dashed lines 
indicate the magnitude range over which the color histogram in the 
upper panel has been constructed. The solid line in the upper panel 
is a Gaussian with $\sigma_{obs} = 0.182$ mag fitted to this histogram. Given 
that the photometric error in this magnitude range is 
$\sigma_{err} = 0.049$ mag as yielded by the artificial stars, this 
implies an intrinsic color dispersion of  $\sigma_{int} = 0.175$ mag 
represented by the dashed Gaussian in the upper panel.}

\figcaption{The lower panel shows our relation between the dereddened 
RGB color at $M_{I}=-2$ [$(V-I)_{o,-2}$] and metal abundance. The 
filled circles are the star clusters listed in Table 2 while the 
dashed curve is the weighted least squares fit shown in Equation 1. 
In light of the photometric error yielded by the artificial stars of
$\sigma_{err} = 0.049$ mag, the upper panel displays the metallicity 
uncertainty resulting purely from the photometric errors.}

\figcaption{Using the dereddened colors of all stars with magnitudes 
in the range $22.55 \leq I \leq 22.75$ ($-1.9 \leq M_{I} \leq -2.1$) 
along with Equation 1, we contruct the metallicity distribution 
function shown in the upper panel by the filled symbols. The solid line is the 
generalized histogram of these abundances, which takes into account 
the metallicity error of each star. The lower panel shows the 
variation of the MDF as the adopted reddening is varied from E(B--V) = 0.06 
(dashed line) to 0.08 (solid line) to 0.10 (dotted line).}

\figcaption{(a) A comparison of the `G272' field metallicity 
distribution function (MDF) and that of the G302 field from Holland et al. 
(1996) located at a projected distance of 32 arcmin from the center of 
M31. (b) Same as (a) except that the MDF of the M31 halo field from
Durrell et al. (1994) located at 40 arcmin from the nucleus is shown. 
(c) Same as (a) except that the
MDF of the M31 halo field located 
at 90 arcmin from the nucleus and studied 
by Durrell et al. (2001) is shown. All of these distributions have 
been scaled to have the same area. (d) The MDFs have been scaled to 
the match the metal-poor tail of the `G272' MDF.}

\figcaption{Gaussian fits to our metallicity distribution function 
after adopting the Durrell et al. (2001) parameters for the 
metal-poor and intermediate metallicity halo components. The upper 
and lower panels show the fits to the binned (filled circles) and 
generalized (solid line) histograms, respectively. The dashed lines 
are the individual Gaussian
components while the dotted line is the sum of all three.}

\figcaption{(a) A comparison of the G272 field metallicity 
distribution function (MDF) with that of Milky Way thick disk stars 
from Wyse \& Gilmore (1995). (b) Same as (a) except that the MDF of 
Milky Way field halo stars from Ryan \& Norris (1991) is shown. 
(c) Same as (a) except that the MDF of Milky Way Globular clusters 
constructed from the database of Harris (1996) is shown. (d) Same as 
(a) except that the MDF of M31 globular clusters from Barmby et al. 
(2000) is shown. All of these distributions have 
been scaled to have the same area.}

\figcaption{(a) The left panel shows the distance and reddening 
corrected $[M_V,(V-I)_{o}]$ color-magnitude diagram (CMD) for Region 1; 
the right panel shows the $[M_I,(V-I)_{o}]$ CMD. The solid lines are 
the Girardi et al. (2000) solar abundance (Z=0.019) theoretical isochrones 
for ages of $10^{8}$, $6.3 \times 10^{8}$, $10^{9}$, and 
$1.6 \times 10^{9}$ years.
(b) Same as (a) except that the Region 2 CMDs are shown. (c) Same as (a) 
except that the Region 3 CMDs are shown.}


\begin{thebibliography}

\bibitem[]{} Alves, D., \& Sarajedini, A. 1999, \apj, 511, 225
    
\bibitem[]{} Barmby, P., Huchra, J. P., Brodie, J. P., Forbes, D. A.,
Schroder, L. L., \& Grillmair, C. J. 2000, \aj, 119, 727

\bibitem[]{} Beaulieu, S., Gilmore, G., Elson, R. A. W., Johnson, R. 
A., Santiago, B., Sigurdsson, S., \& Tanvir, N., 2001, astro-ph/0102312

\bibitem[]{} Bellazzini, M., Ferraro, F., \& Pancino, E. 2001, ApJ, in 
press (astro-ph/0104114)

\bibitem[]{} Bica, E., Barbuy, B., \& Ortolani, S. 1991, \apj, 382, L15

\bibitem[]{} Burstein, D., \& Heiles, C. 1982, \aj, 87, 1165

\bibitem[]{} Bullock, J. 1999, PhD Thesis, University of California,
Santa Cruz

\bibitem[]{} Caputo, F., Castellani, V., Chieffi, A., Pulone, L., \& 
Tornamb\'{e}, A. 1989, \apj, 340, 241

\bibitem[]{} Castellani, V., Chieffi, A., \& Pulone, L. 1991, \apjs, 
76, 911

\bibitem[]{} Chaboyer, B., Green, E. M., \& Liebert, J. 1999, \aj, 117, 1360

\bibitem[]{} Cole, A. 1998, \apj, 500, L137

\bibitem[]{} Crowl, H. H., Sarajedini, A., Piatti, A., Geisler, D., 
Bica, E., Clar\'{a}, J. J., \& Santos, J. F. C. 2001, \aj, 122, 220

\bibitem[]{} Cuillandre, J. -C., Lequeux, J., Allen, R. J., Mellier, 
Y., \& Bertin. E. 2001, \apj, in press (astro-ph/0102350)

\bibitem[]{} Da Costa, G. S., \& Armandroff, T. E. 1990, \aj, 100, 162

\bibitem[]{} Da Costa, G. S., Armandroff, T. E., Caldwell, N., \&
Seitzer, P. 2000, \aj, 119, 705

\bibitem[]{} Davidge, T. J. 1993, \apj, 409, 190

\bibitem[]{} Durrell, P. R., Harris, W. E., \& Pritchet, C. J. 1994, 
\aj, 108, 2114

\bibitem[]{} Durrell, P. R., Harris, W. E., \& Pritchet, C. J. 2001, 
astro-ph/0101436

\bibitem[]{} Ferraro, F. 1992, Mem. Soc. Astron. Italiana, 63, 491

\bibitem[]{} Ferraro, F. R., Messineo, M., Fusi Pecci, F., 
de Palo, M. A., Straniero, O., Chieffi, A., \& Limongi, M. 1999, \aj, 
118, 1738

\bibitem[]{} Fusi Pecci, F., Ferraro, F. R., Crocker, D. A., 
Rood, R. T., \& Buonanno, R. 1990, \aap, 238, 95

\bibitem[]{} Fusi Pecci, F., Bellazzini, M., Cacciari, C., \& Ferraro,
F. R. 1995, \aj, 110, 1664

\bibitem[]{} Fusi Pecci, F., Buonanno, R., Cacciari, C., Corsi, C. E.,
Djorgovski, S. G., Federici, L., Ferraro, F. R., Parmeggiani, G.,
\& Rich, R. M. 1996, \aj, 112, 1461

\bibitem[]{} Gallart, C. 1998, \apj, 495, L43

\bibitem[]{} Garnavich, P. M., VandenBerg, D. A., Zurek, D. R.,
\& Hesser, J. E. 1994, \aj, 107, 1097

\bibitem[]{} Girardi, L., \& Salaris, M. 2001, \mnras, in press

\bibitem[]{} Girardi, L., Bressan, A., Bertelli, G., \& Chiosi, C.,
2000, \aaps, 141, 371

\bibitem[]{} Gould, A. 1994, \apj, 435, 573

\bibitem[]{} Grebel, E. K. 2000, in A New Era of Microlensing 
Astrophysics, ASP Conf. Ser., edited by J. W. Menzies \& P. D. Sackett,
in press (astro-ph/0008249)

\bibitem[]{} Guarnieri, M. D., Ortolani, S., Montegriffo, P., Renzini, 
A., Barbuy, B., Bica, E., \& Moneti, A. 1998, \aap, 331, 70

\bibitem[]{} Harris, W. E. 1996, 
    http://physun.mcmaster.ca/$\sim$harris/mwgc.dat
    
\bibitem[]{} Hodge, P., \& Lee, M. -G. 1988, \apj, 329, 651

\bibitem[]{} Hodge, P., Lee, M. -G., \& Mateo, M. 1988, \apj, 324, 
172

\bibitem[]{} Holland, S., Fahlman, G., \& Richer, H. 1996, \aj, 112, 1035

\bibitem[]{} Holland, S., Côté, P., \& Hesser, J. E. 1999, \aap, 348, 418

\bibitem[]{} Jablonka, P., Bridges, T., Sarajedini, A., Meylan, G., 
Maeder, A., \& Meynet, G. 1999, \apj, 518, 627

\bibitem[]{} Jablonka, P., Courbin, F., Meylan, G., Sarajedini, A., 
Bridges, T. J., \& Magain, P. 2000, \aap, 359, 131

\bibitem[]{} Johnson, J., \& Bolte, M. 1998, \aj, 115, 693

\bibitem[]{} Layden, A. C., \& Sarajedini, A. 1997, \apj, 486, L107

\bibitem[]{} Lee, Y. -W., Demarque, P., \& Zinn, R. J. 1990, \apj, 
350, 155

\bibitem[]{} Lee, Y. -W., Demarque, P., \& Zinn, R. J. 1994, \apj, 
423, 248

\bibitem[]{} Massey, P., Armandroff, T. E., \& Conti, P. S. 1986, \aj, 
92, 1303

\bibitem[]{} Morris, P. W., Reid, I. N., Griffiths, W. K., \& Penny, 
A. J. 1994, \mnras, 271, 852

\bibitem[]{} Mould, J. R., \& Kristian, J. 1986, \apj, 305, 591

\bibitem[]{} Ortolani, S., Barbuy, B., \& Bica, E. 1990, \aap, 236, 362

\bibitem[]{} Pritchet, C. J., \& van den Bergh, S. 1988, \apj, 331, 135

\bibitem[]{} Renzini, A. 1999, in When and How do Bulges Form and 
Evolve?, edited by C. M. Carollo, H. C. Ferguson, \& R. F. G. Wyse 
(Cambridge:Cambridge University Press), p. 9

\bibitem[]{} Rich, R. M., Mighell, K., \& Neill, J. D. 1996, in Formation
of the Galactic Halo...Inside and Out, ASP Conf. Ser. Vol. 92, edited
by H. Morrison \& A. Sarajedini (ASP:San Francisco) p. 544

\bibitem[]{} Richer, H. B., Crabtree, D. R., \& Pritchet, C. J. 1990, 
\apj, 355, 448

\bibitem[]{} Ryan, S. G., \& Norris, J. E., 1991, \aj, 101, 1865

\bibitem[]{} Sakai, S. Zaritsky, D., \& Kennicutt, R. C., Jr. 2000,
\aj, 119, 1197

\bibitem[]{} Sarajedini, A. 1994, \aj, 107, 618

\bibitem[]{} Sarajedini, A. 1999, \aj, 118, 2321

\bibitem[]{} Sarajedini, A., \& Forrester, W. L. 1995, \aj, 109, 1112

\bibitem[]{} Sarajedini, A., Lee, Y. -W., \& Lee, D. -H. 1995, \apj, 
450, 712

\bibitem[]{} Sarajedini, A., Geisler, D., Schommer, R., \& Harding, P.
2000, AJ, 120, 2437

\bibitem[]{} Saviane, I., Rosenberg, A., Piotto, G., \& Aparicio, A. 
2000, \aap, 355, 966

\bibitem[]{} Schlegel, D. J., Finkbeiner, D. P., \& Davis, M. 1998,
\apj, 500, 525

\bibitem[]{} Silbermann, N. A. et al. 1996, \apj, 470, 1

\bibitem[]{} Silk, J. 2000, \mnras, in press (astro-ph/0010624)

\bibitem[]{} Stanek, K. Z., \& Garnavich, P. M. 1998, \apj, 503, 131

\bibitem[]{} Stetson, P. B. 1994, \pasp, 106, 250

\bibitem[]{} van den Bergh, S. 1999, \araa, 9, 273

\bibitem[]{} Walker, A. 1994, \aj, 108, 555

\bibitem[]{} Williams, B. F., \& Hodge, P. 2001, ApJ, 548, 190

\bibitem[]{} Wyse, R., \& Gilmore, G., 1995, \aj, 110, 2771

\bibitem[]{} Zoccali, M., Renzini, A., Ortolani, S., Bica, E., \& 
Barbuy, B. 2001, astro-ph/0101200

\end{thebibliography}
\end{document}